# Wide range screening of algorithmic bias in word embedding models using large sentiment lexicons reveals underreported bias types

**Authors:** David Rozado[1]*

**Affiliations**

[1]Otago Polytechnic, New Zealand.

*Correspondence to: drozado@gmail.com

**Abstract:** Concerns about gender bias in word embedding models have captured substantial attention in the algorithmic bias research literature. Other bias types however have received lesser amounts of scrutiny. This work describes a large-scale analysis of sentiment associations in popular word embedding models along the lines of gender and ethnicity but also along the less frequently studied dimensions of socioeconomic status, age, physical appearance, sexual orientation, religious sentiment and political leanings. Consistent with previous scholarly literature, this work has found systemic bias against given names popular among African-Americans in most embedding models examined. Gender bias in embedding models however appears to be multifaceted and often reversed in polarity to what has been regularly reported. Interestingly, using the common operationalization of the term *bias* in the fairness literature, novel types of so far unreported bias types in word embedding models have also been identified. Specifically, the popular embedding models analyzed here display negative biases against middle and working-class socioeconomic status, male children, senior citizens, plain physical appearance and intellectual phenomena such as Islamic religious faith, non-religiosity and conservative political orientation. Reasons for the paradoxical underreporting of these bias types in the relevant literature are probably manifold but widely held blind spots when searching for algorithmic bias and a lack of widespread technical jargon to unambiguously describe a variety of algorithmic associations could conceivably be playing a role. The causal origins for the multiplicity of loaded associations attached to distinct demographic groups within embedding models are often unclear but the heterogeneity of said associations and their potential multifactorial roots raises doubts about the validity of grouping them all under the umbrella term *bias*. Richer and more fine-grained terminology as well as a more comprehensive exploration of the bias landscape could help the fairness epistemic community to characterize and neutralize algorithmic discrimination more efficiently.

## Introduction

The term *algorithmic bias* is often used to imply systematic offsets in algorithmic output that produce unfair outcomes such as privileging or discriminating an arbitrary group of people over others. The topic of algorithmic bias has recently elicited widespread attention among the artificial intelligence research community. Popular machine learning artifacts have been used to illustrate the creeping of purported societal biases and prejudices into models used for computer vision (1), recidivism prediction (2) or language modeling (3).

Word embedding models are dense vector representations of words learned from a corpus of natural language (4). Word embeddings have revolutionized natural language processing due to their ability to model semantic similarity and relatedness among pairs of words as well as linear regularities between words that roughly capture meaningful cultural constructs such as gender or social class (5). The usage of word embeddings in upstream natural language processing tasks has often improved the accuracy of those systems downstream (6).

Word embedding models have been claimed to capture prejudicial bias, stemming from the corpus on which they were trained, against women and ethnic minorities (7). Indeed, it has been shown that popular word embedding models tend to associate word vector representations of popular given names among African Americans with negative terms (8), female given names with words such as *nursing* and *homemaker* and male given names with prestigious professions such as *computer programmer* and *doctor* (3).

This work has examined the existing literature on the creeping of societal biases into word embedding models through a systematic search of the sources ArXiv, dblp Computer Science Bibliography, Google Scholar and Semantic Scholar for the queries: *word embeddings bias* and *word vectors bias*. The queries were carried out on April 26, 2019. A total of 28 papers were identified where the Abstract clearly addresses the topic of bias in word embeddings (see Table 1 in the Appendix for details). Examination of the Abstracts and Introduction sections of the manuscripts revealed that 26 (93%) of them cited the issue of gender bias and 15 (54%) cited bias along racial or ethnic lines. Other types of biases such as those due to age or religiosity were only marginally mentioned (10%) or not at all like bias due to political or sexual orientation. Of the 26 papers addressing gender bias, 19 (73%) specifically described gender bias detrimental to females and none considered the possibility of gender bias in word embeddings detrimental to males.

The overwhelming focus of the existing literature on the topic of gender bias, the consistency of the reported bias direction and the lack of attention paid to other bias types, such as viewpoint biases, motivated this work to carry out a systematic analysis of a wide range of possible biases potentially creeping into widely used word embedding models.

This work systematically analyzed 3 popular word embeddings methods: Word2vec (Skip-gram) (4), Glove (9) and FastText (10), externally pretrained on a wide array of corpora such as Google News, Wikipedia, Twitter or Common Crawl. The ability of each model to capture semantic similarity, relatedness as well as morphological, lexical, encyclopedic and lexicographic analogies (11) was measured (see Table 1). FastText models slightly outperformed Word2vec and Glove models probably due to their ability to model morphological relationships at the subword level.

Word embeddings have been proven remarkably capable of capturing valid quantitative associations about the empirical world by encoding such information in the geometrical arrangement of words in vector space. To test for the existence of associations in word embeddings, previous works have often measured the cosine similarity between two sets of words (7,8). More recently, other authors have proposed deriving from the embedding space cultural axes representing constructs such as gender or race and then projecting words of interests onto those axes to test for associations (12). Both approaches yield analogous results since they are algebraically similar. In this work, we use cultural axes to measure associations due to their more intuitive interpretation.

Cultural axes are created by subtracting an aggregate of related words representing one end of a spectrum (Pole 1) from another set of opposite words representing the other end of the spectrum (Pole 2) (3,5). Hence, a gender axis can be created by subtracting a male pole formed by aggregating a basket of archetype word vectors representing males such as *male*, *man* and *men* from a female pole derived from an aggregate of word vectors representing females such as *female*, *woman* and *women* (Figure 9). Any word vector in the model vocabulary can then be projected onto the gender axis. The landing location of the projection on the gender axis can be used to test whether the model tends to associate said word, representing for instance a profession such as *lawyer*, with the male or the female pole of the gender axis (Figure 10). If there is a systematic association of a set of words denoting high status professions with one gender, the model is claimed to have bias (8,13). But projecting words onto cultural axes can also reveal that word vectors encode a surprising amount of associations, other than bias, about the empirical world.

Figure 1 (A) illustrates, as shown in previous works (8), that creating a gender axis in the Google News Word2vec embedding space associates words denoting professions that have a large percentage of female representation with the female pole of the gender axis. Conversely, professions with a low percentage of female representation are associated with the male pole. That is, the value of word vectors representing professions projected onto a gender axis correlates significantly with the percentage of female representation in said professions. Creating an economic development axis, see Figure 1 (B), or a price axis, see Figure 1 (C), associates rich countries and expensive car manufacturer brands with the prosperous poles of the axes and poor countries and affordable car manufacturer brands with the impoverished poles. The same figure also shows that the projection of words representing professions onto a political orientation axis, see Figure 1 (D), mildly correlates with empirical data on political campaign donations by those professional groups.

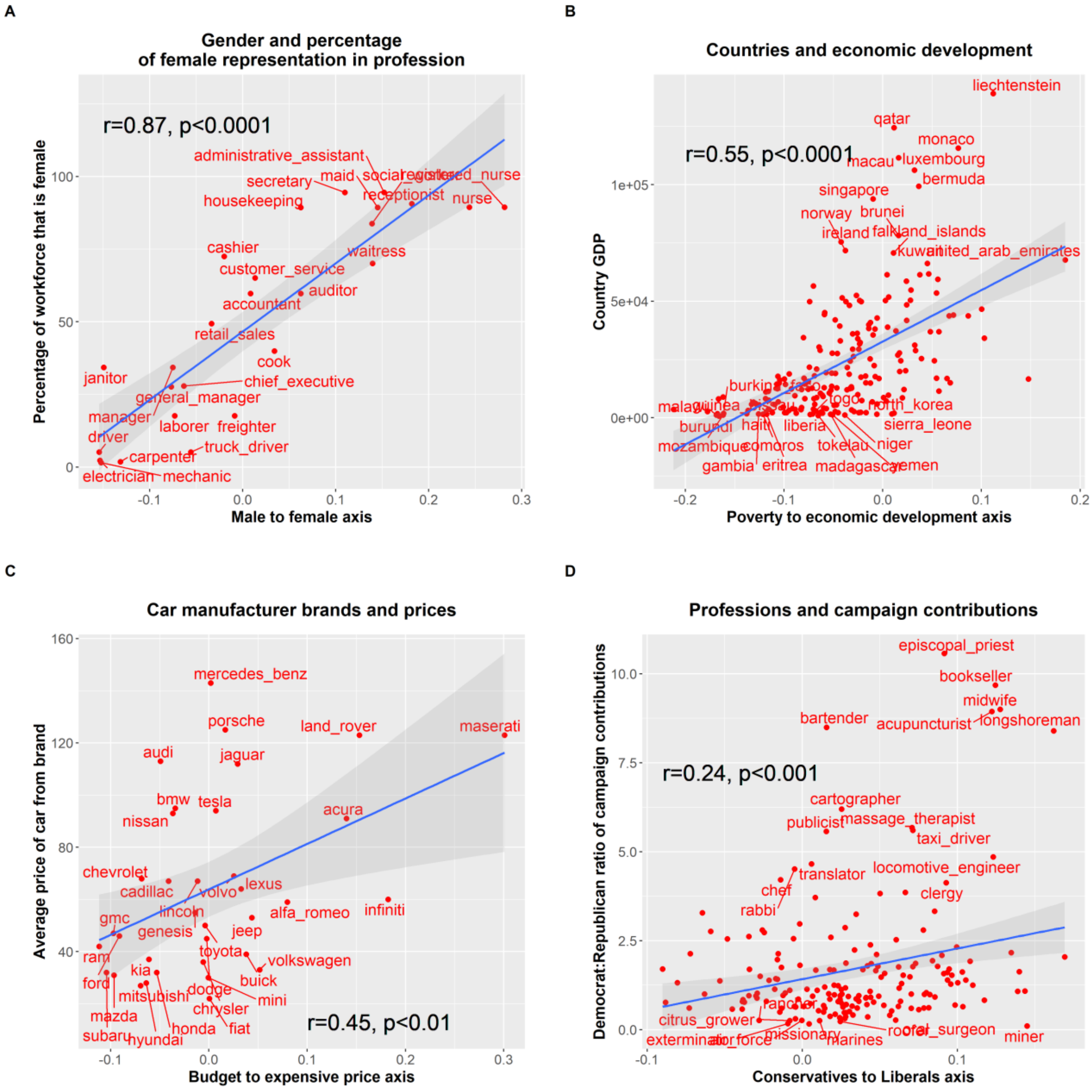

**Figure 1 Plenty of associational knowledge about the world is encoded in the geometrical structure of a word2vec model trained on Google News. This information can be retrieved by projecting vector representations of target words onto cultural axes. The figure shows significant correlation (Pearson) between word vectors describing professions projected onto a gender axis and the percentage of female representation in those professions (A). Similarly, the value of projecting word vectors of countries onto an economic development axis correlates with the GDP of the country (B). Car brands word vectors projected onto a price axis also correlate with the average price of cars from a given brand (C). Even word vectors representing professions projected onto a political orientation axes correlate, albeit mildly, with the Democrat:Republican ratio of political campaign contributions within the profession (D). Details about the data sources used in the Y axes of the figure are provided in the Methods section.**

Previous works have often applied the aforementioned or similar methodology to test for associations in word embedding models (often conceptualized as biases) between small sentiment lexicons, such as WEAT (8) (N=50), and a reduced set of demographic groups according to ethnicity (8) or gender (13). This work proposes the usage of larger lexicons of positive/negative terms for a more systemic scrutiny of biases entrenched within popular embedding models. The main analysis of this work is carried out using the Harvard General Inquirer IV-4 Positiv-Negativ lexicon (HGI) (14), containing 3623 unique manually labeled positive and negative terms, that has been widely used in the content analysis literature. We also replicate the experiments using 16 additional sentiment lexicons, including WEAT, to demonstrate that most associations reported herein are not circumscribed to the HGI lexicon.

The most noteworthy contribution of this work has been to systematically test popular pre-trained word embedding models for the existence of a wide array of possible biases beyond gender and ethnicity. This is done by creating cultural axes describing demographic categories along the lines of gender and ethnicity but also along the lines of sexual orientation, religiosity, age, socioeconomic status, physical appearance and political opinion. Details about axis creation and the words forming their poles are provided in the Methods section and the Appendix. Words from the HGI sentiment lexicon, where each entry has a positive or negative label, are then projected onto the cultural axes. The null hypothesis is that there is no correlation between sentiment words annotations (positive or negative) and their projection values on a cultural axis. That is, that the embedding model does not preferentially associate positive/negative terms with neither pole of a cultural axis.

A significant positive or negative correlation between sentiment lexicon labels and the projection values of the sentiment lexicon onto a cultural axis reflects a preferential association of positive/negative terms with a distinct pole of the axis. If the pole of the axis represents a specific demographic group, this is interpreted as bias to maintain consistency with the prevailing usage of the term in previous scholarly work that used similar methodology to test for ethnic or gender bias (8,13). In subsequent analysis, we use the nonparametric Spearman rank correlation coefficient that makes few assumptions about the distribution of the data. A positive correlation coefficient denotes an association of positive sentiment terms with Pole 2 of the axis and a corresponding association of negative terms with Pole 1. A negative correlation denotes an association of positive terms with Pole 1 and a corresponding association of negative terms with Pole 2.

To validate the methodology proposed in this work, four illustrative cultural axes are created in the Google News word2vec model using poles with widely accepted positive/negative connotations. The four axes are: death-life, disease-health, dictatorship-democracy and malevolent historical figures-respectable historical figures. Details about the words composing the axes poles are provided in the Appendix. Projecting the HGI lexicon onto these axes results in significant correlations between the sentiment labels and the projection values of the sentiment words on the axes analyzed. That is, positive terms in the HGI lexicon (labeled as +1) tend to be associated with the axes positive range representing life, health, democracy and respectable historical figures. Conversely, negative terms in the HGI lexicon (labeled as -1) tend to be associated with the axes negative range representing death, disease, dictatorship and malevolent historical figures, see Figure 2.

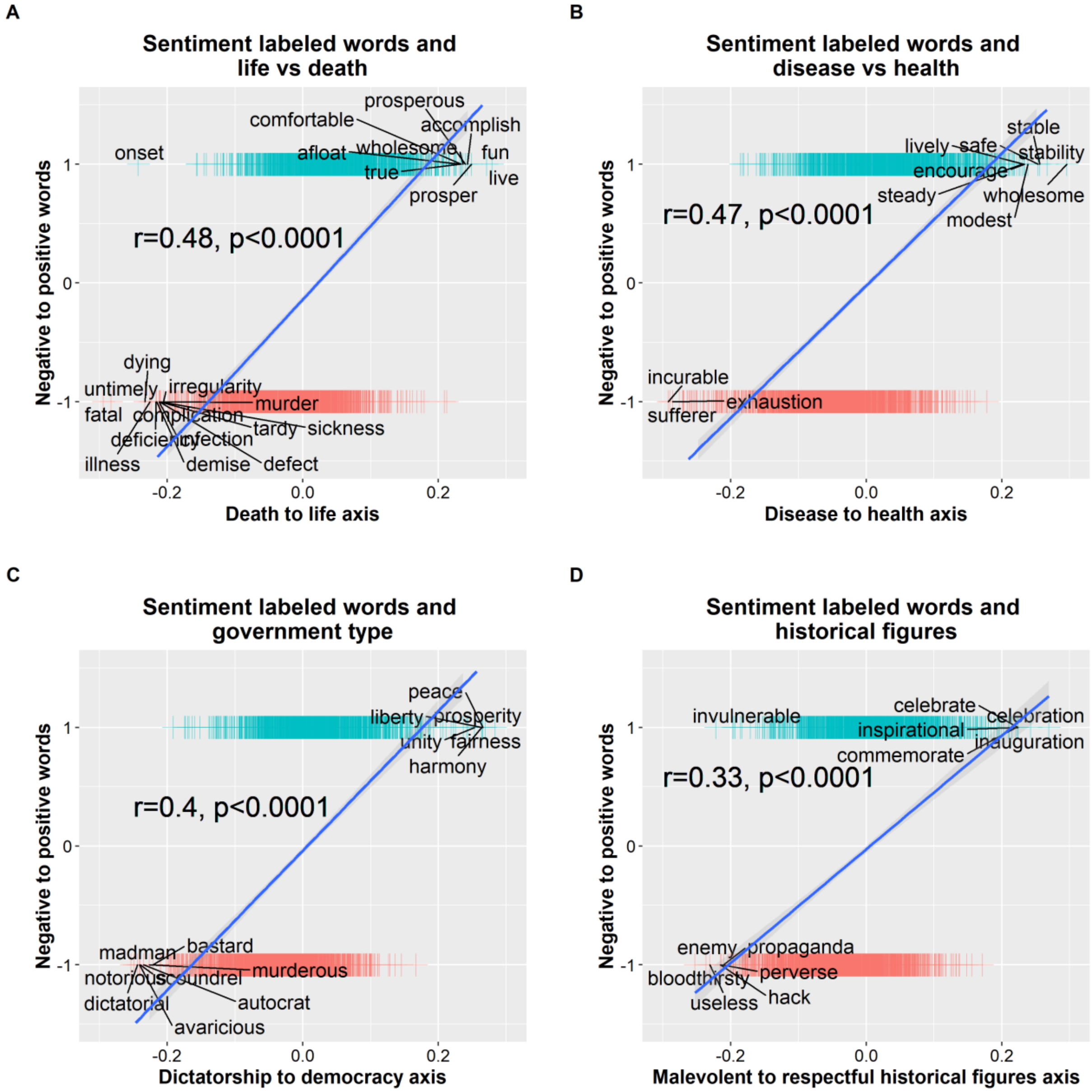

**Figure 2 To validate the methodology used in subsequent analysis, this figure shows that systematically projecting a large sentiment lexicon (HGI) on cultural axes with widespread agreed-upon positive/negative polarity results in positive terms in the lexicon being associated with the poles representing life, health, democracy and respectful historical figures. Conversely, negative terms tend to be associated with the poles representing death, disease, dictatorship and malevolent historical figures.**

## Results

Figure 3 shows the results of projecting the HGI lexicon onto the cultural axes denoting demographic groups that are the focus of this work. Detailed numerical results of all correlations for each axis and model as well as the corresponding p-values with respect to the null hypothesis of no preferential association of positive terms with neither pole of the cultural axes are provided in Table 2 of the Appendix. P-values were adjusted for multiple comparisons using the conservative Bonferroni method but they are often extremely small due to the large sample size of the HGI lexicon.

Most popular pretrained word embedding models tend to preferentially associate positive lexicon terms with the feminine poles of most gender axes and conversely negative terms with the masculine poles. Notably, this is not the case for the axis formed using popular masculine and feminine given names where the association is close to neutral in 5 of the 7 embedding models analyzed. The only embedding model displaying an apparent lack of overall gender bias is the Glove model trained on the Twitter corpus.

The results for the ethnic axes are mixed. Some embedding models display a heterogeneity of bias directions and effect sizes for general words used to describe ethnic groups (such as Whites, Hispanics, Asians or African-Americans) while others display neutral associations. Most embedding models however tend to associate negative terms with given names popular among African-Americans.

Youth, good looks and high socio-economic status are consistently associated with positive terms and conversely, the elderly, plain looking and those of working or middle-class socioeconomic status are associated with negative terms. Results for the sexual orientation axis are not conclusive with different embedding models displaying neutral or opposite bias directions. Several embedding models display negative bias against lack of religious faith. In the Christians to Muslims axis, there is a consistent mild to moderate association of negative terms with the pole representing Muslims. Finally, there is a pervasive association of conservative individuals and ideas with negative terms and an association of liberals with positive terms along all the popular pre-trained embedding models analyzed.

The results of the association experiments were highly correlated between the different embedding models analyzed (average correlation of 0.77, see Table 3 in the Appendix for details). This suggests that distinct embedding models are independently capturing similar latent associations that are widely common within the different corpora on which the models were trained.

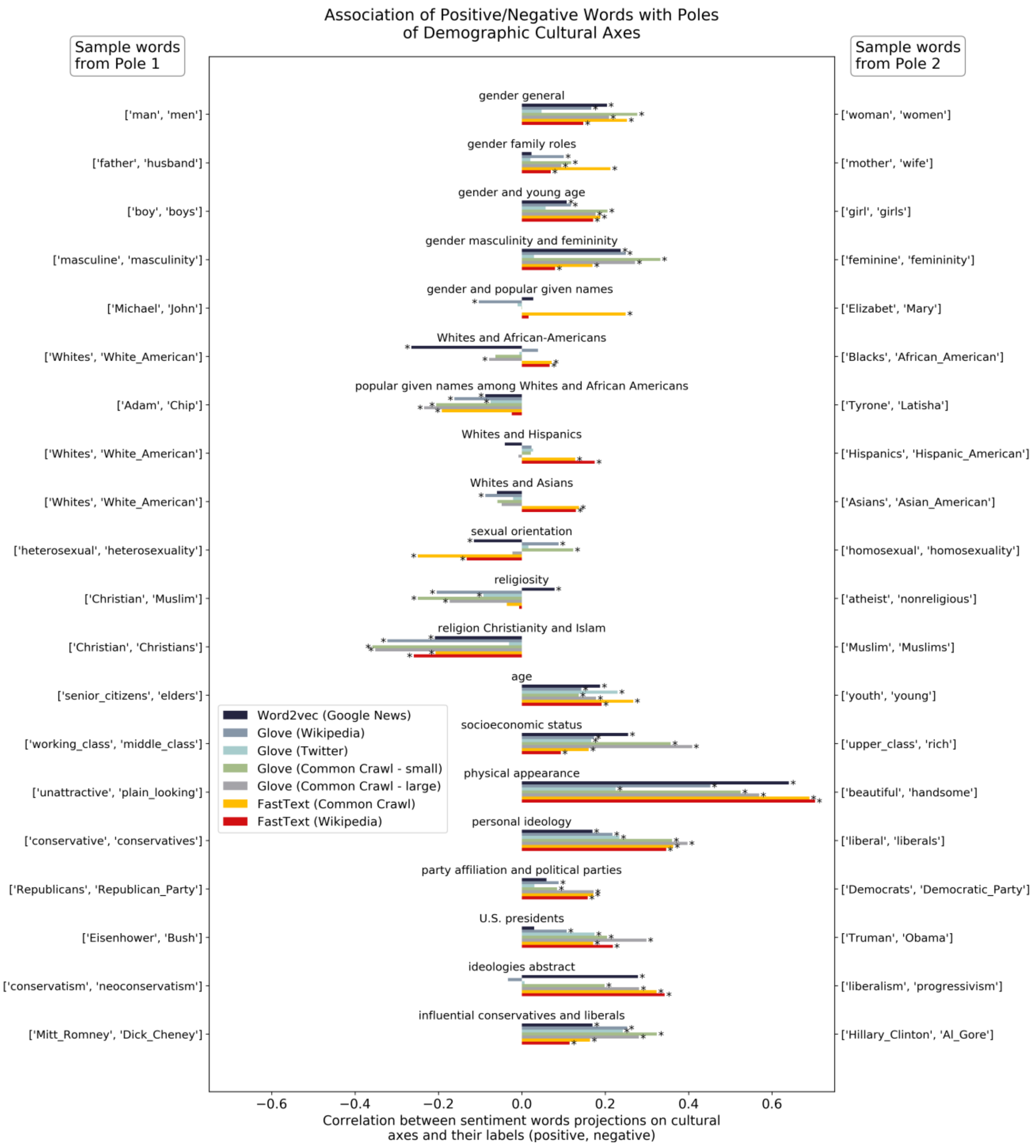

**Figure 3 Projecting a large sentiment lexicon (HGI, N=3623) with positive/negative labeled terms on a set of cultural axes along the lines of gender, ethnicity, age, physical appearance, socioeconomic status and political/sexual/religious orientation often shows asymmetrical associations of positive and negative terms with the poles of cultural axes in seven popular pre-trained word embedding models. A positive correlation value denotes association of positive terms with Pole 2 of the cultural axes and conversely association of negative terms with Pole 1. A negative correlation denotes association of positive terms with Pole 1 of the cultural axes and conversely, association of negative terms with Pole 2. Asterisks adjacent to a model bar indicate a statistically significant result (p<0.01) after applying the Bonferroni adjustment for multiple comparisons.**

A reasonable criticism of the methodology used in this and related papers to test for bias in embedding models is that the choice of terms forming the poles of the cultural axes does not follow a systematic approach, which could raise doubts about the universality of the results obtained. However, due to the geometrical structure of embedding spaces, by which similar terms tend to be located in adjacent locations of vector space, adding or removing words with similar semantic meaning from a pole construct should not have a major effect on axis orientation. After all, adding two normalized and similar vectors to each other will result in a location in vector space that is similar in orientation to either vector. This entails that the vector resulting from adding the word vectors *beautiful* and *handsome* will be very similar in orientation to the vector resulting from the addition of the word vectors *beautiful*, *handsome* and *pretty*. Obviously, adding a completely unrelated term to a pole will have a larger impact on axis orientation. Also, poles containing many terms should have more stable axes orientations since addition or removal of a single term will have only a minor effect on the aggregate of the remaining vectors.

To test these assumptions, we excised in turns random samples of 25%, 50% and 75% of the words composing the poles of a cultural axis, used the remaining terms to re-create the axis, reran the experiment shown in Figure 3 500 times and averaged the results across models and repetitions for each excision size. The resulting averaged correlation coefficients are shown in Figure 4. For axes with an excision size of 25% of the original terms forming the poles, results are almost indistinguishable to the original axes formed with full poles. Larger excisions sizes result in mild alterations of axes orientation and often smaller effect sizes but rarely collapse to a reversal of bias polarity. This suggests that cultural axes orientation are robust to moderate excision or addition of semantically related terms. As expected, poles that contain many terms (such as male/female given names, or popular given names among African-Americans/Whites) are particularly robust to removal of even 75% of the original terms forming the poles.

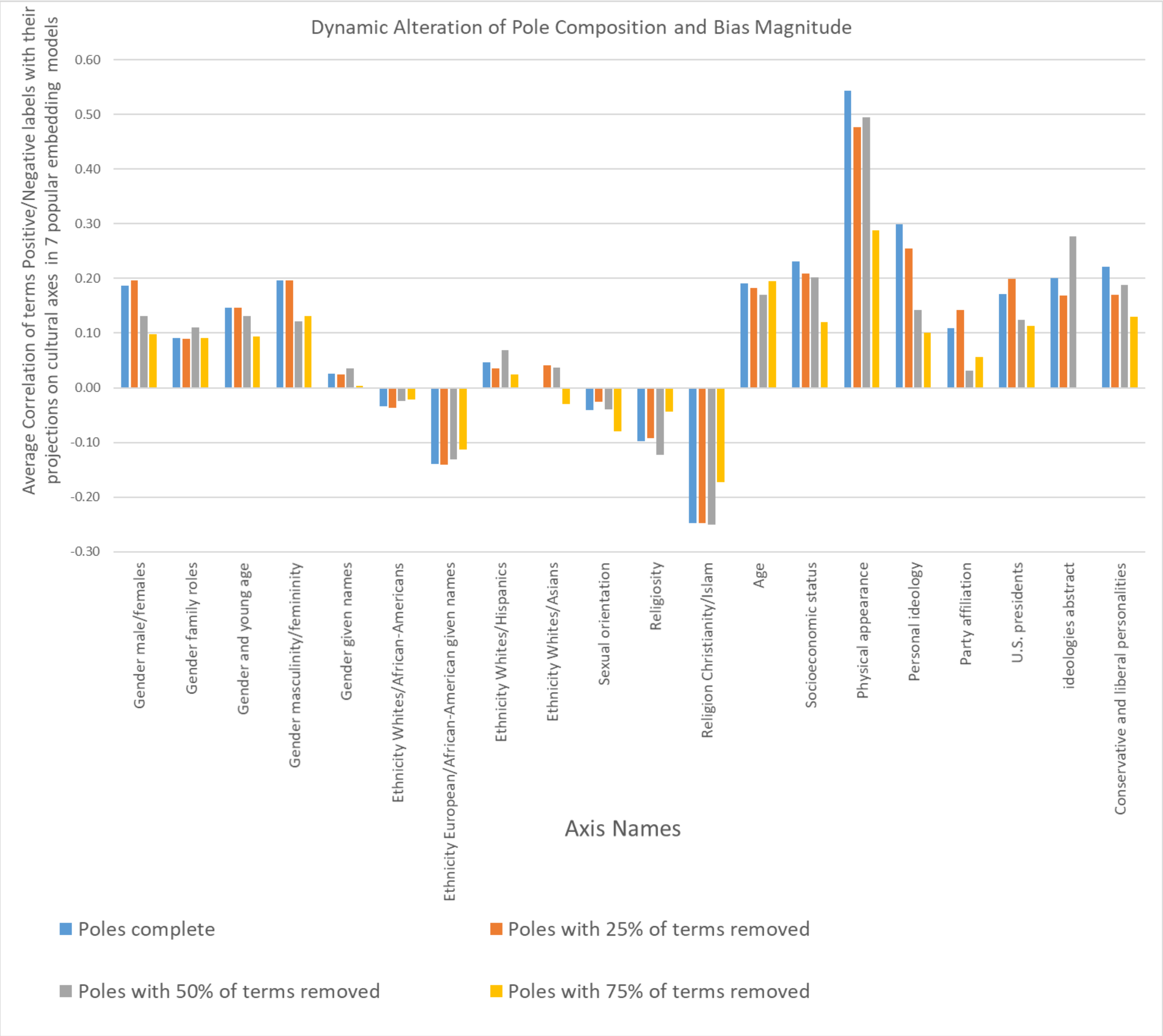

**Figure 4 Results of altering the composition of words forming the poles of cultural axes. Blue bars represent the average bias across the 7 word embedding models analyzed for a given cultural axis. Orange bars show bias magnitudes for the same cultural axes with random samples of 25% of the terms forming its poles excised. Gray and yellow bars show the results of the association experiments when 50% and 75% of the words forming the poles are removed respectively. Bar heights for different excision sizes are the result of running each random experiment 500 times and averaging the results across repetitions and embedding models.**

In order to obtain conclusive proof that the degree and direction of bias present in popular word embedding models is not an artifact of the HGI lexicon, an ensemble of 16 additional external lexicons pre-annotated for positive and negative terms was used to probe for bias in embedding models (see Table 4 in the Appendix). Many of these lexicons have been widely used in the sentiment and content analysis research literature. Systematically projecting different lexicons in the cultural axes analyzed generated results similar to those obtained with the HGI lexicon. Bias magnitudes were highly correlated regardless of sentiment lexicon used to test for bias (average correlation of 0.84, see Table 5 in the Appendix), suggesting that all lexicons are measuring a similar construct of negativity and positivity. Note that some of these extra lexicons have non-binary sentiment annotations, so results in Figure 3 are not circumscribed only to lexicons with binary labels.

The mild association of positive words with the feminine poles of most gender axes persisted in the ensemble analysis. The association of negative terms with the pole constructed with African-American given names was also replicated. There was a very mild but consistent association of negative terms with common nouns used to refer to African-Americans for most of the lexicons tested. It is important to point out that this trend was not apparent when using the HGI lexicon alone to test for bias along this axis. On aggregate, associations for terms used to describe other ethnic minorities such as Hispanics and Asians were mostly heterogeneous and small or close to neutral.  Again, the association of negative terms with the pole representing Muslims in the Christians – Muslims axis was replicated with all the additional lexicons. Similarly, there was also a very mild association of lack of religiosity with negative terms. Finally, there was a mild to moderate association of negative terms with the conservative poles of the political orientation axes regardless of sentiment lexicons used. As in previous experiments, associations of positive terms with youth, beauty and wealth persisted.

One of the additional lexicons tested, the WEAT lexicon, deserves special consideration since previous works have used this small size lexicon (N=50) when testing for bias in word embedding models (8,15,16). Although results of projecting WEAT sentiment words onto the cultural axes analyzed roughly agree with the HGI lexicon projection tests, some cultural axes show divergent results. For example, in the male to females gender axis (row 1 in Table 4 of the Appendix), the WEAT lexicon suggests that there is no association of positive/negative terms with neither pole of the axis. This is contrarian to all the other lexicons tested that often show a tendency to have their positive terms associated with the female pole of the gender axis. Similarly, in the religiosity axis, the WEAT lexicon hints at the existence of a negative bias against religious faith while most of the other sentiment lexicons show a mild negative bias against the opposite pole: lack of religious sentiment. These results suggest that the WEAT small set size can sometimes result in misleading results when trying to measure systemic bias in word embedding models. Thus, larger lexicons, with a wider coverage of a language vocabulary, are likely to be more robust in detecting systemic bias than smaller lexicons.

Nonetheless, even a single large lexicon such as HGI can occasionally deviate from the consensus results emerging out of an ensemble analysis that tests for latent associations using several sentiment lexicons. For example, the HGI lexicon seems to show, on aggregate, a lack of substantial bias against common nouns used to refer to African-Americans across embedding models. Yet, many other lexicons often show a very subtle but consistent negative bias against these terms. This hints at the multifaceted nature of bias and the occasional inadequacy of a single lexicon, even if voluminous in size, to fully characterize the entire bias space.

To visually illustrate the association of sentiment words with the poles of cultural axes in word embedding models, Figure 5 shows the entire vocabulary contained in the 17 sentiment lexicons used in this work (N=15,635) projected onto a political orientation axis and a gender axis on the 7 popular word embedding models analyzed. Positive and negative labeled words are color-coded blue and red respectively for ease of visualization. A clear trend to associate positive words with femininity and liberals and in turn negative terms with masculinity and conservatives is apparent across the word embedding models analyzed. Positive words mostly dominate the upper right quadrant of the plane, denoting femininity and liberal political orientation and negative words dominate the lower left quadrant, denoting masculinity and conservatives.

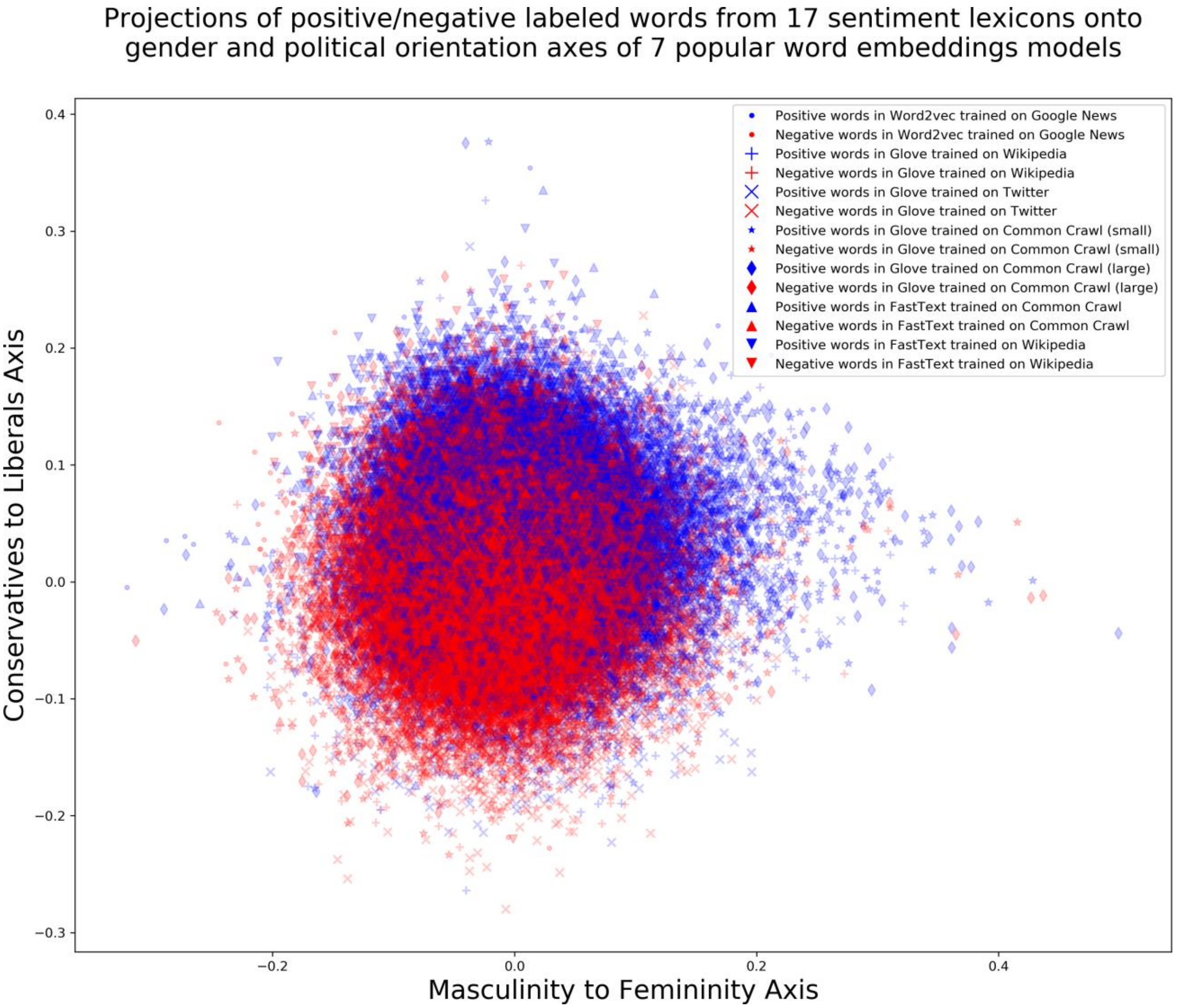

**Figure 5. Associations of externally annotated positive/negative terms ($N = 15635$; $negative = 9181$) from 17 sentiment lexicons with gender and political orientation cultural axes derived from 7 popular word embedding models. Blue color denotes a positive word and red color denotes a negative word.**

To demonstrate that the results in Figure 5 are not dominated by a single large lexicon, Figure 6 shows that the associations of positive/negative terms with distinct poles of the cultural axes are prevalent across lexicons and embedding models (see also detailed Tables 2 & 4 in the Appendix).

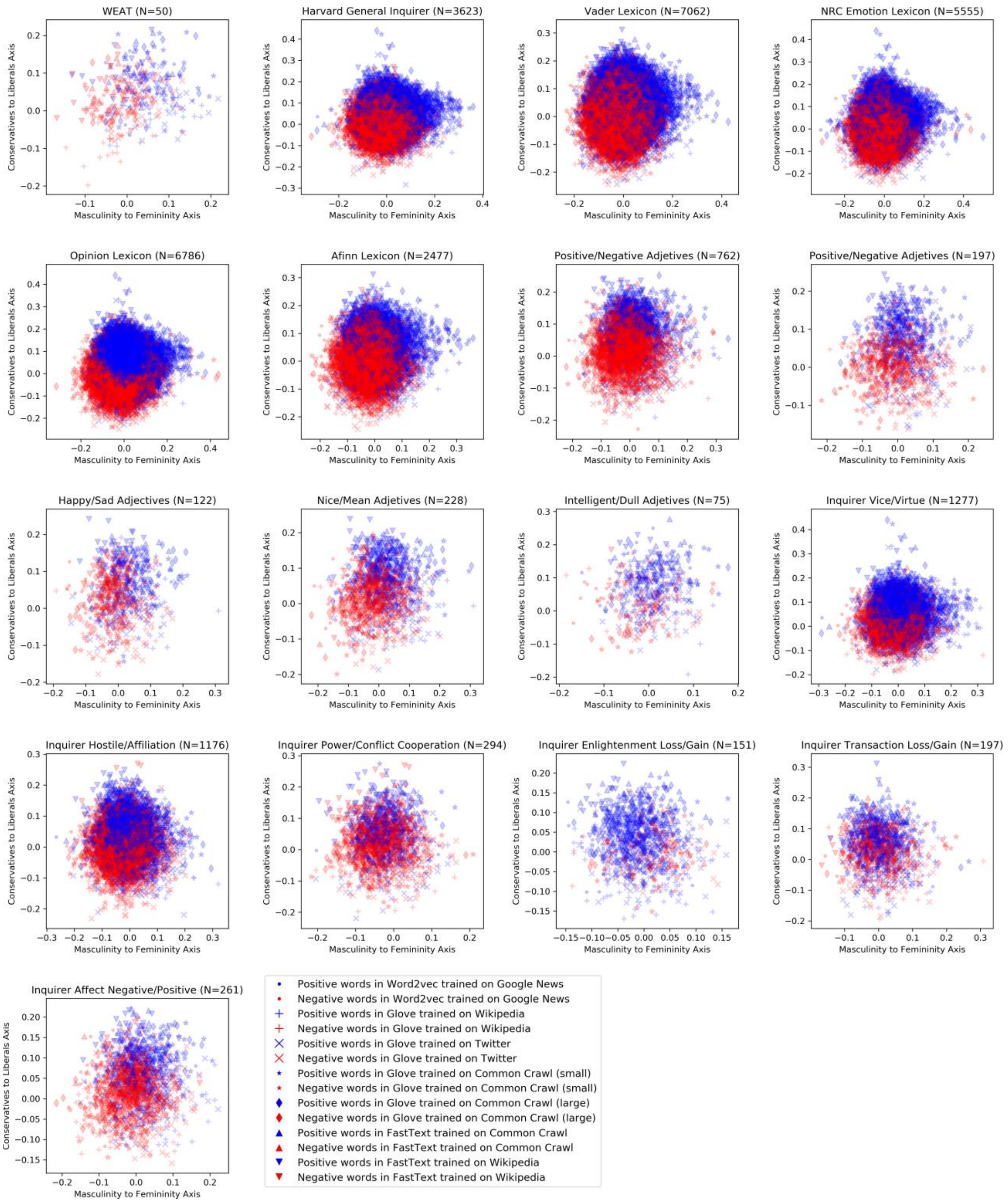

**Figure 6. Projections of words in 17 sentiment lexicons onto gender and political orientation axes for the 7 word embedding models analyzed. Positive words are color-coded in blue and negative words are color-coded in red.**

A final consideration about the results reported above is that the label "negativity" in sentiment lexicons sometimes conflates victimization and reprehensible behavior or character traits. That is, both the words *sufferer* and *uncompassionate* are often labeled as negative in sentiment lexicons. Yet, the word *sufferer* denotes being at the receiving end of victimization and can elicit empathy towards the subject characterized as such. In contrast, the word *uncompassionate* suggests a character flaw and can elicit hostility against the subject labelled as such. Therefore, a natural question about the negative latent associations embedded in language models towards a given demographic group is whether said negative associations denote empathy towards a victim by emphasizing someone's suffering and victimization or whether these associations portray an individual or group as exhibiting negative character traits or behavior.

It is possible to visualize the specific nature and geometry of the strongest positive/negative associations in embedding models by comparing orientation similarity between the cultural axes denoting demographic groups that are the focus of this work and axes built using antonym pairs retrieved from Wordnet (N=3872), see Figure 12 for methodological details. That is, an antonym pair from WordNet such as *unintelligent-intelligent* can be used to trace the direction in embedding space moving from lack of cognitive ability towards intelligence. The cosine similarity between this vector and a cultural axis such as political orientation can be estimated to quantify the degree of alignment between each word in the antonym pair and the poles of the cultural axis.

Figure 1 in the Appendix shows the results of comparing the orientation of the personal ideology cultural axis with 3872 axes formed using WordNet antonym pairs across the seven popular embedding models studied. Only the top 30 most similar WordNet antonym pairs derived axes are shown. A clear tendency to associate conservatives with negative character traits is apparent in most embedding models. Pole 1 (representing conservatives) in the personal ideology cultural axis is often aligned with words such as regressive, narrowminded, unenlightened, unsupportive, undemocratic, inhuman, unprofessional, uncompassionate, intolerant, impolite, stingy, uneducated, intolerant, uncooperative, unintelligent, sectarian, prejudiced and violent. These are all words that denote negative character traits rather than emphasizing someone's suffering. In contrast, Pole 2 (representing liberals) is associated with words such as broad-minded, enlightenment, supportive, democratic, edifying, generous, dignified, compassionate, humane, generous, cooperative, unprejudiced, nonviolent and tolerant. Similar Figures ranking the similarity of axes derived from Wordnet antonym pairs with gender, ethnicity, and religious cultural axes are also provided in the Appendix.

Figure 1 in the Appendix can be challenging to visually digest and contains many Wordnet antonym pairs without an obvious positive/negative valence. To aid with the visualization of specific sentiment polarity across embedding models with regard to political orientation, Figure 7 displays words from the top 100 alignments between axes derived from Wordnet antonym pairs and the conservatives to liberal axis, with the constraint that antonym pairs need to also belong to the HGI sentiment lexicon. That is, the words appearing in Figure 7 have been externally labeled for positive/negative polarity and belong to the antonym pairs that better align with the political orientation cultural axis. Words are color-coded blue and red according to positive/negative polarity. A clear trend to align Wordnet negative antonyms with the conservative pole and positive antonyms with the liberal pole is apparent across all embedding models. Similar Figures of Wordnet sentiment antonyms projected onto gender, ethnicity, age, physical appearance, socioeconomic and religious cultural axes are provided in the Appendix.

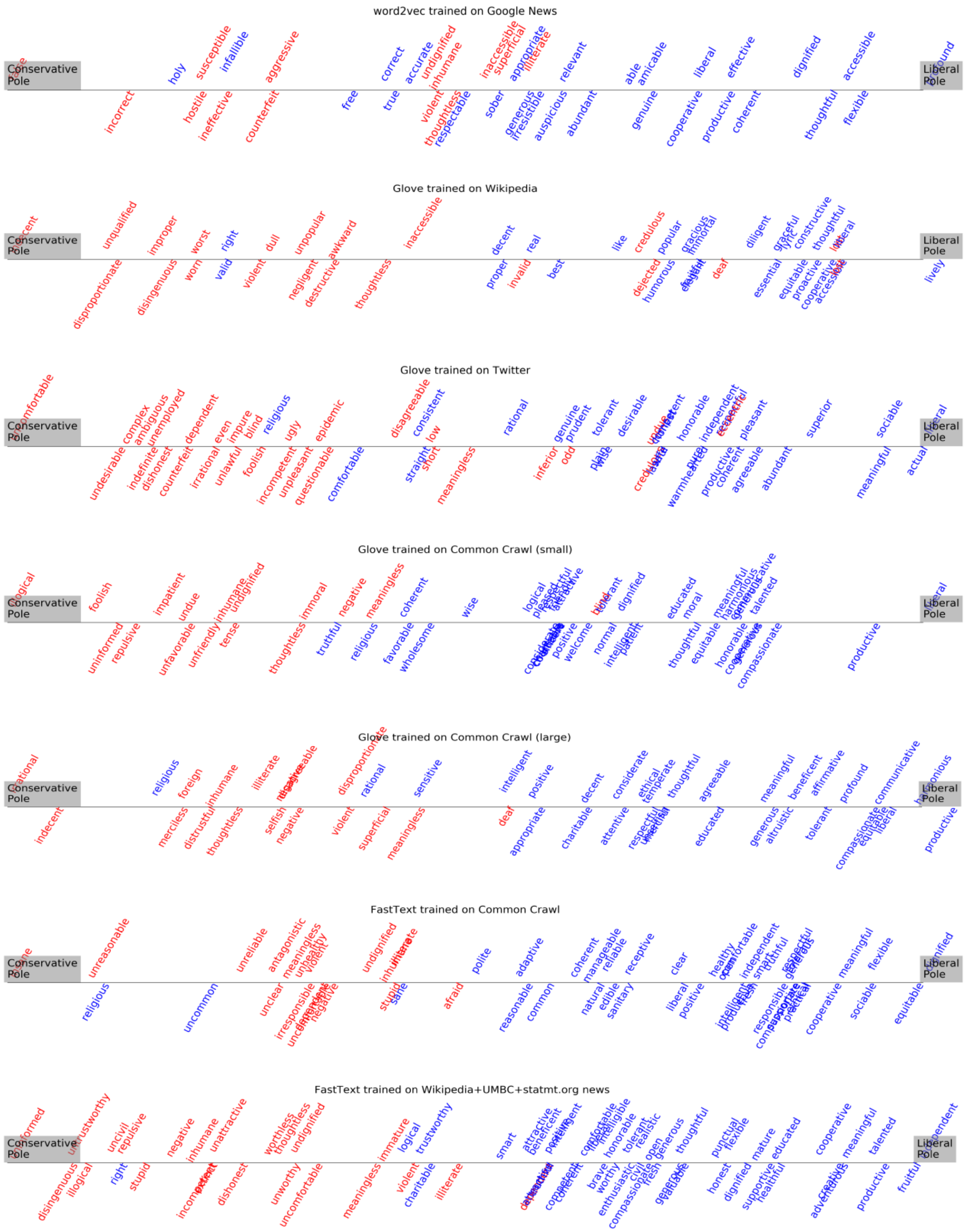

**Figure 7 HGI sentiment lexicon words among Wordnet top 100 antonym pairs derived axes that better align in orientation with the conservatives to liberals cultural axis in 7 popular pre-trained embedding models. Words have been color-coded red and blue to signify negative and positive labels. The word 'progressive' has been left out from the figure since it was an outlier in several axes that distorted the visualization.**

## Discussion

The most noteworthy result of this work is the finding in most word embedding models of significant associations between negative sentiment words and terms used to denote conservative individuals or ideas, people of middle and working-class socioeconomic status, underage and adult males, senior citizens, Muslims, lack of religious faith, and given names popular among African-Americans. Albeit, the later bias had already been reported previously (8). These results suggest the existence of systemic bias, as the term has been used in the existing literature, in most widely used word embedding models against the aforementioned groups. This is relevant since embedding models are routinely used as subcomponents of larger automated systems that profile individuals in social media networks and other digital systems. Furthermore, the fact that similar biases are apparent on models that were trained using a variety of different corpora (Wikipedia, Google News, Twitter or Common Crawl) suggests that these associations are common across a wide range of texts.

The results presented here also suggest that the widespread concerns in the algorithmic fairness research literature on the issue of gender bias in word embeddings models and in particular, biases that are detrimental to females, are perhaps overstated. Although biased associations between female denoting terms and certain words certainly do exist across embedding models, the comprehensive analysis provided herein reveals that, on aggregate, word embedding models tend to preferentially associate terms that denote females with positive sentiment words and terms that denote males with negative sentiment words (see Figure 3).

Obviously, focusing any analysis on a reduced set of terms can only provide a partial and incomplete impression about word embeddings biases detrimental to a particular group. Previous works have sometimes centered their analysis on a narrow set of terms, for instance specific professions, to conclude that gender bias in embedding models exists (3). Words representing certain prestigious professions such as *programmer* or *engineer* are indeed closer to words denoting males than to words denoting females in most word embedding models, as widely reported in the literature (3,8,13), but so are the less prestigious and underreported words *janitor*, *plumber*, *beggar* or *murderer*. This suggests that the bias landscape is multifaceted and that while particular associations of a subset of words with a human population group can be conceptualized as a specific bias type, only a comprehensive analysis of a diverse set of large lexicons manually annotated for sentiment polarity can more precisely characterize the existence, or lack thereof, of *systemic* bias against human groups in language models.

On a related note, the occasional contradictory results between association analyses using the popular WEAT lexicon, containing just 50 terms, and the consensus results of the ensemble analysis using several larger sentiment lexicons carried out in this work, suggests that the WEAT lexicon can be sometimes limiting when trying to detect systemic bias in word embeddings. This is probably due to the small set size of WEAT. Thus, larger lexicons are better suited to provide a more comprehensive overview of embedding models tendencies to systematically associate human groups with positive or negative terms. But even larger lexicons can sometimes be insufficient to fully characterize the multidimensional nature of the bias space. This was illustrated in the Results section by the inability of the HGI lexicon to detect, on aggregate, a slight bias across most embedding models against common nouns used to refer to African-Americans. Only the ensemble analysis using several sentiment lexicons was able to provide hints about the existence of this very subtle bias. Thus, sentiment lexicons, even large ones, can only probe specific regions of the bias space and their results do not necessarily generalize to other regions of the bias landscape.

The consistent associations of negative terms with old age, middle or working-class socioeconomic status and below average physical appearance reported in this work deserves special consideration. Whether these associations should be operationalized as bias is perhaps debatable. On the one hand, humans dread and avoid death (17) which is strongly correlated with advanced age, so it is perhaps understandable that negative terms are associated with the pole denoting old age. Alternatively, some societies highly value and revere their elders as custodians of wisdom and tradition (18). Perhaps in those cultures, the association of negative terms with words describing senior citizens is not apparent. Similar arguments can be made about wealth and beauty. Most humans have a tendency to seek those attributes and avoid their antonyms (19,20). The key question is when do the resulting negative associations creeping into machine learning models exceed being mere psychological or cultural human predispositions and become stereotypes and instruments of discrimination that marginalize specific subgroups of the population. Thus, operationalizing a precise definition of the term *bias* and charting to which demographic populations or under which circumstances it can be applied is not straightforward. To compound the difficulty, the usage of the term *bias* in the relevant literature has displayed substantial semantic elasticity, making its precise characterization even more elusive.

It is a sociological fact that there exists an unequal distribution of gender representation in different occupations of the US labor market. Word embedding models that absorb that statistical reality and tend to associate words such as *man* or *men* with male-dominated professions such as *plumbing* or *engineering* and terms such as *woman* or *women* with female dominated professions such as *nursing* or *midwifery* are often conceptualized in the contemporary machine learning literature as biased (2, 3, 8). Furthermore, machine learning models that predict recidivism risk on an equal treatment basis at the individual level but that do not generate equal outcome rates across population groups are also being deemed as biased (21). The implications of this semantic elasticity and its broad applicability makes precise communication of research results around the topic of algorithmic bias challenging.

Using a common modern operationalization of the term *bias* in the algorithmic bias literature with respect to gender or ethnicity, that is, deviations from equal algorithmic outcomes at the group level for distinct demographic groups, irrespective of underlying population distributions, denote bias, it is an inescapable conclusion of this work's results that the popular pre-trained word embedding models analyzed here contain biases along the lines of gender and ethnicity, but also along the lines of political/religious orientation, age, physical appearance and socioeconomic status, due to the unequal distribution of positive and negative lexicon terms associations along the poles of the aforementioned cultural axes. That is, sentiment lexicons projections onto cultural axes representing demographic groups deviate from the equal outcomes benchmark, irrespective of group baseline rates, that has often been established as desirable for other non-uniform algorithmic outputs.

Notwithstanding the claims above, the author has concerns about the frequent elastic usage of the term *bias* to describe a heterogeneous array of distinct algorithmic phenomena. The lack of precise and widespread rich terminology to refer to a variety of algorithmic behavior embedded within machine learning models, and that are often simply grouped under the umbrella term bias, makes conceptualization, operationalization and communication of research results about algorithmic fairness unnecessarily confusing.

Perhaps it would be helpful if distinct and specific terms beyond *bias* would be routinely used to refer to a variety of machine learning models behavior when classifying or characterizing human groups. That is, a specific term could be consistently used in the literature to refer to models that simply capture empirically valid but nonuniform statistical structure across population groups. A different term could be used to refer to the very severe discriminatory case of algorithmic outputs that have disparate impact on protected and unprotected groups due to algorithmic failure to properly model the statistical structure of the underlying features driving the classification. Additional terms could also be consistently used to describe machine learning models that treat subjects equally at the individual level versus models that disregard equal treatment in order to enforce equality of outcome at the group level. Richer and more precise terminology would provide fine-grained characterization and interpretation of algorithmic output. This in turn would help the field to communicate more efficiently. Rigorous efforts to outline the different interpretations of fairness by introducing precise terminology exist (22) but the usage of such fine-grained vocabulary is not yet apparent in the survey of the word embeddings bias literature carried out by this work.

The novel results conveyed herein regarding underreported algorithmic bias types also highlight the scant attention paid by the algorithmic bias research literature to biases in word embedding models due to socioeconomic status, political orientation, age, and religious faith. The analysis of the existing research literature failed to reveal a single paper that has previously reported the existence of bias against ideological viewpoints in word embedding models. This work cannot provide conclusive proof about the reasons for the paradoxical underreporting of the biases novelly reported in this work. Conceivably, researchers in the fairness community could simply be following the trend of pioneer work, as it happens in other scientific disciplines, that serendipitously just happened to focus on gender biases in word embeddings. Alternatively, the preferential exploration of certain bias types in embedding algorithms could reflect a blind spot bias within the algorithmic bias epistemic community that manifests itself as a predilection for exploring certain regions of the research landscape, such as gender bias against females, while avoiding other regions such as biases against middle or working-class socioeconomic status, conservative individuals, senior citizens, Muslims or male children (i.e. boys). This bias might emerge in part from the viewpoint composition of the Academy were most researchers are located or where they spent their formative years.

It is well established that most elite research universities and liberal arts colleges lack viewpoint diversity along moral and political orientation axis, with large Democrat to Republican ratios existing across most faculty departments (23,24). A liberal ideological orientation is markedly sensitive to concerns around discrimination against females and ethnic minorities (25). It is also well-established that cognitive biases influence scientists' choosing of research interests, interpretation of research results and reception of research outcomes (26). This could partly explain the overwhelming interest in the word embeddings research literature for gender bias but the sidelining of other bias types. Although plausible, this hypothesis fails to explain the lack of attention in the embedding models' literature to other underreported bias types, such as those against religious minorities, that perhaps should have also attracted the interest of a liberal Academy (see Table 1 in the Appendix). Thus, additional mechanisms are likely to be at play as well, such as embedding models extraordinary ability to capture gender associations with high degree of fidelity (see Figure 1A very large correlation coefficient) and their less stellar performance in capturing more complex or subtle associations such as socioeconomic status or political orientation (see Figure 1C or Figure 1D more modest correlation coefficients).

Given the importance of accurately detecting, scrutinizing and addressing algorithmic bias, an important question that derives from this work is how to minimize potential blind spot biases within the fairness epistemic community that itself scrutinizes algorithmic bias. If this research community is susceptible to widely held community blind spots, proper characterization and neutralization of a comprehensive set of algorithmic biases could remain elusive. Fine grained technical jargon to precisely describe algorithmic behavior, comprehensive and open-ended explorations of the bias landscape and adversarial collaboration within intellectually heterodox working groups could serve as instruments that minimize community blind spots.

The author of this work is almost certainly infused with biases himself. Yet, juxtaposition of empirically backed heterodox viewpoints, even if partially biased, can help members of an epistemic community to mitigate each other's blind spots. It is in that spirit that this work has been put forward. All the materials necessary to reproduce the results reported herein are provided (https://github.com/drozado/WideRangeScreeningOfAlgorithmicBiasInWordEmbeddings).

## Methods

Word embeddings

Word embeddings are a set of language modeling and feature learning techniques used in natural language processing (NLP) for mapping words in a corpus vocabulary to dense vector representations *(5)*. Embedding models use the distributional statistics of human language to capture the semantic and syntactic roles of a word in a given corpus (see Figure 8). Thus, word vectors are positioned in vector space such that words that share common contexts in the corpus are located in close proximity to one another in vector space. The embeddings also capture regularities in vector space such as constant vector offsets between related words that usually convey culturally meaningful connotations such as gender or socioeconomic status. Popular methods to generate word embeddings from a corpus of natural language include neural networks and dimensionality reduction on the word co-occurrence matrix. This work analyzed bias in three popular word embeddings algorithms: Word2vec, Glove and FastText.

Word2vec *(4)* is a group of related architectures (Continuous Bag of Words or CBOW and Skip-gram) consisting of a shallow, two-layer neural network trained to reconstruct the linguistic contexts of words. GloVe *(9)* is a global log-bilinear regression model that combines the advantages of the two major model families in the embeddings literature: global matrix factorization and local context window methods. Both Word2vec and Glove ignore the morphology of words, by assigning a distinct vector representation to each word. This is limiting for languages with large vocabularies and many rare words. FastText (10) overcomes this limitation by extending the Word2vec Skip-gram model to represent each word as a bag of character n-grams. A vector is associated to each character n-gram and individual words are represented as the sum of these individual n-grams. For example, the vector representation of the word *mouse* is obtained by adding the n-grams vectors "*<mo*", "*mou*", "*mous*", "*mouse*", "*mouse>*", "*ous*", "*ouse*", "*ouse>*", "*use*", "*use>*", "*se>*" assuming hyperparameters of smallest

n-gram=3 and largest ngram=6. This method allows FastText to compute word representations for out of vocabulary words (words that did not appear in the training corpus).

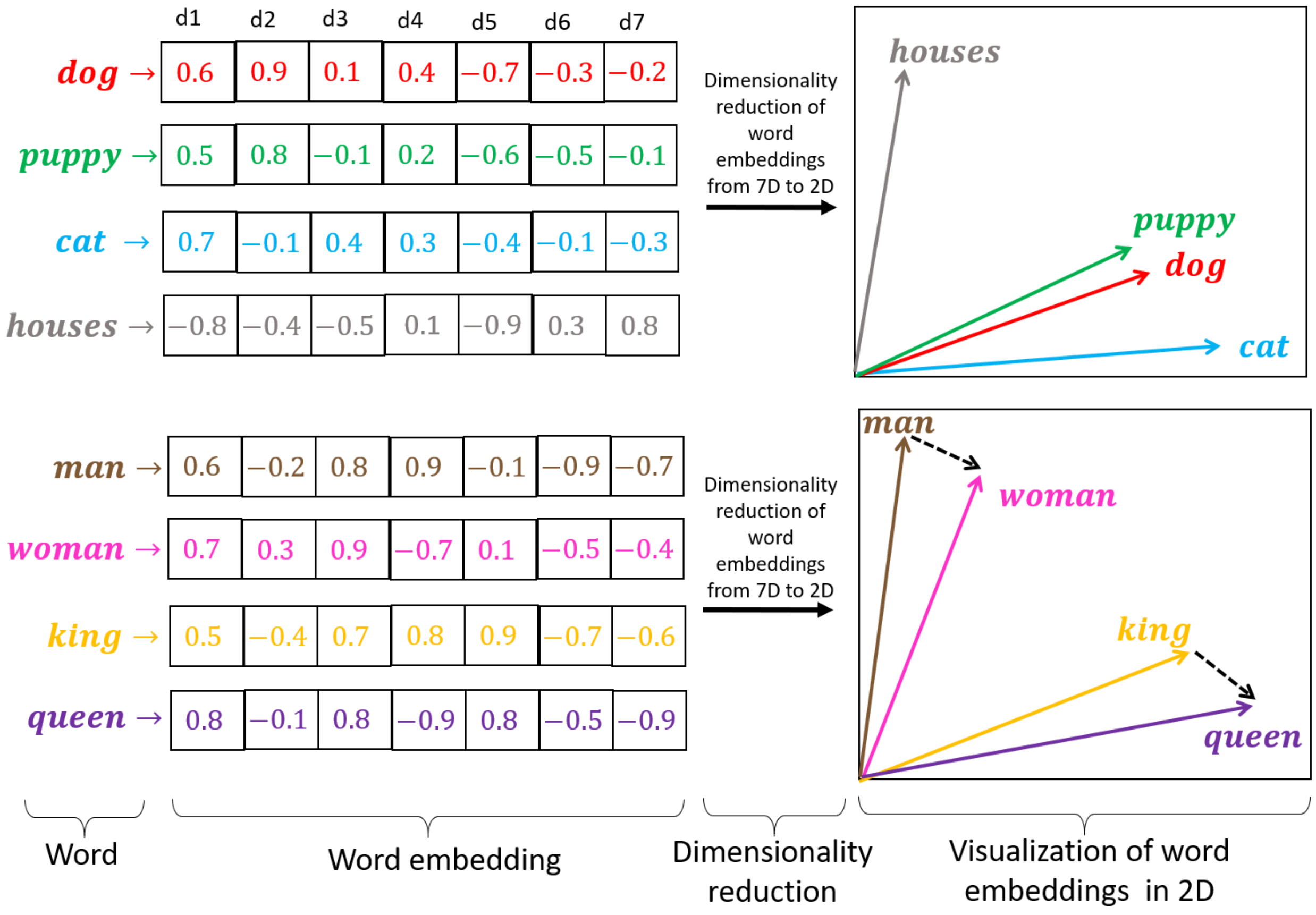

**Figure 8. Word embeddings map words in a corpus of text to vector space. Linear combinations of dimensions in vector space correlate with the semantic and syntactic roles of the words in the corpus. For illustration purposes, dimension *d1* in the figure has a high positive correlation with living beings. A properly tuned word embedding model will map words with similar semantic or syntactic roles to adjacent regions in vector space. This property can be visualized through dimensionality reduction techniques such as t-SNE or PCA (see upper right quadrant of the figure). Cultural concepts are also apparent in vector space as consistent offsets between vector representations of words sharing a particular relationship. For instance, in the bottom right of the figure, the dotted vector represents a gender regularity that goes from masculinity to femininity.**

## Survey of the computer science literature on the topic of bias in word embeddings models

A search of the computer science literature using the engines ArXiv (https://arxiv.org), DBLP Computer Science Bibliography (https://dblp.uni-trier.de), Google Scholar (https://scholar.google.com/) and Semantic Scholar (https://www.semanticscholar.org) for the queries *word embeddings bias* and *word vectors bias* identified 28 papers with a focus on the topic of bias in word embeddings models as inferred from the Abstract. Manuscripts were classified according to the bias types that they cite (gender, race, etc.) by examining their Title, Abstract and Introduction sections. A tabular listing of the manuscript titles and the bias types that they mention is provided in Table 1 of the Appendix.

Word embedding models analyzed

Seven popular and readily available Word2vec, Glove and FastText models pre-trained on different types of corpora were used for the analysis of biases in word embeddings. The seven word embeddings models analyzed and the corpora on which they were externally trained are listed below:

-Word2vec Skip-Gram trained on Google News corpus (100B tokens)
https://code.google.com/archive/p/word2vec/

-Glove trained on Wikipedia 2014 + Gigaword 5 (6B tokens)
http://nlp.stanford.edu/data/glove.6B.zip

-Glove trained on 2B tweets Twitter corpus (27B tokens)
http://nlp.stanford.edu/data/glove.twitter.27B.zip

-Glove trained on Common Crawl small (42B tokens)
http://nlp.stanford.edu/data/glove.42B.300d.zip

-Glove trained on Common Crawl large (840B tokens)
http://nlp.stanford.edu/data/glove.840B.300d.zip

-Fastext trained with subword information on Common Crawl (600B tokens)
https://dl.fbaipublicfiles.com/fasttext/vectors-english/crawl-300d-2M-subword.zip

-FastText trained with subword information on Wikipedia 2017, UMBC webbase corpus and statmt.org news dataset (16B tokens)
https://dl.fbaipublicfiles.com/fasttext/vectors-english/wiki-news-300d-1M-subword.vec.zip

Model evaluation

In order to assess the quality of different word embeddings models, the ability of each model to assess word pairs similarity and relatedness as well as morphological, lexical, encyclopedic and lexicographic analogies was measured (see Table 1). Standard validation data sets commonly used in the NLP literature to evaluate the quality of word embeddings were used.

| Model index | 1 | 2 | 3 | 4 | 5 | 6 | 7 |
|---|---|---|---|---|---|---|---|
| Word embedding algorithm | Word2vec (Skip-Gram) | Glove | Glove | Glove | Glove | FastText | FastText |
| Vector dimensions | 300 | 300 | 200 | 300 | 300 | 300 | 300 |
| Training corpus name | Google News | Wikipedia + Gigaword | Twitter | Common Crawl small | Common Crawl large | Common Crawl | Wikipedia 2017 + UMBC webbase + statmt.org news |
| Corpus size in number of tokens | 100B | 6B | 27B | 42B | 840B | 600B | 16B |
| Model vocabulary size | 3M | 400K | 1.2M | 1.9M | 2.2M | 2M | 2M |
| WordSim-353 | 0.62 | 0.60 | 0.54 | 0.64 | 0.61 | 0.66 | 0.61 |
| MEN similarity dataset | 0.68 | 0.74 | 0.61 | 0.74 | 0.68 | 0.71 | 0.67 |
| SimLex-999 | 0.45 | 0.39 | 0.15 | 0.40 | 0.39 | 0.46 | 0.43 |
| Google Semantic analogies | 0.75 | 0.78 | 0.50 | 0.83 | 0.81 | 0.88 | 0.88 |
| Google Syntactic analogies | 0.74 | 0.67 | 0.60 | 0.69 | 0.74 | 0.84 | 0.90 |
| BATS1 Inflectional Morphology analogies | 0.68 | 0.60 | 0.51 | 0.64 | 0.64 | 0.85 | 0.92 |
| BATS2 Derivational Morphology analogies | 0.17 | 0.09 | 0.08 | 0.14 | 0.18 | 0.32 | 0.42 |
| BATS3 Encyclopedic Semantics analogies | 0.21 | 0.25 | 0.18 | 0.28 | 0.28 | 0.30 | 0.30 |
| BATS4 Lexicographic Semantics analogies | 0.06 | 0.07 | 0.07 | 0.09 | 0.08 | 0.10 | 0.09 |
| **AVERAGE** | **0.48** | **0.47** | **0.36** | **0.49** | **0.49** | **0.57** | **0.58** |

**Table 1 Word embedding models were evaluated using a variety of word similarity, relatedness and analogy tasks often used in the NLP literature. All tests were performed using the top 200,000 most frequent words in each model vocabulary. The BATS test sets sometimes contain several valid answers. We only used the first of those answers to test for the validity of the embedding model predictions.**

Creating cultural axes that trace the spectrum between demographic groups

In a normalized word embedding model, all vectors are unit length. Thus, their semantic and syntactic loading is exclusively determined by vector direction. As described in *(5)*, terms representing similar entities can be aggregated into a construct representative of the group. Figure 9 shows the sum of related vectors for the terms *man* ($v_{man}$) and *men* ($v_{men}$) and subsequent length normalization to create a male vector construct $\hat{v}_M$. An opposing female construct can be created by adding the vectors $v_{woman}$ and $v_{women}$ and normalizing the length of the resulting vector to create a female vector construct $\hat{v}_F$. The substraction $\hat{v}_F - \hat{v}_M$ creates a vector $v_G$ pointing from the male pole $\hat{v}_M$ to the female pole $\hat{v}_F$. Normalizing and centering $v_G$ results in a gender axis $\hat{v}_G$. We can project vector representations of any term in the embedding model vocabulary onto this axis to get a measurement of their degree of association with the male or female poles in the corpus on which the word embedding model was trained. In a word embedding model trained on a sufficiently large corpus containing archetypal cultural associations between professions and gender, the vector representation for the word *midwife* ($v_{midwife}$) will tend to project to the female pole of the gender axis $v_{Gmidwife}$. An archetypal masculine profession such as *priest* ($v_{priest}$) will on the other hand tend to project to the male pole of the gender axis $v_{Gpriest}$.

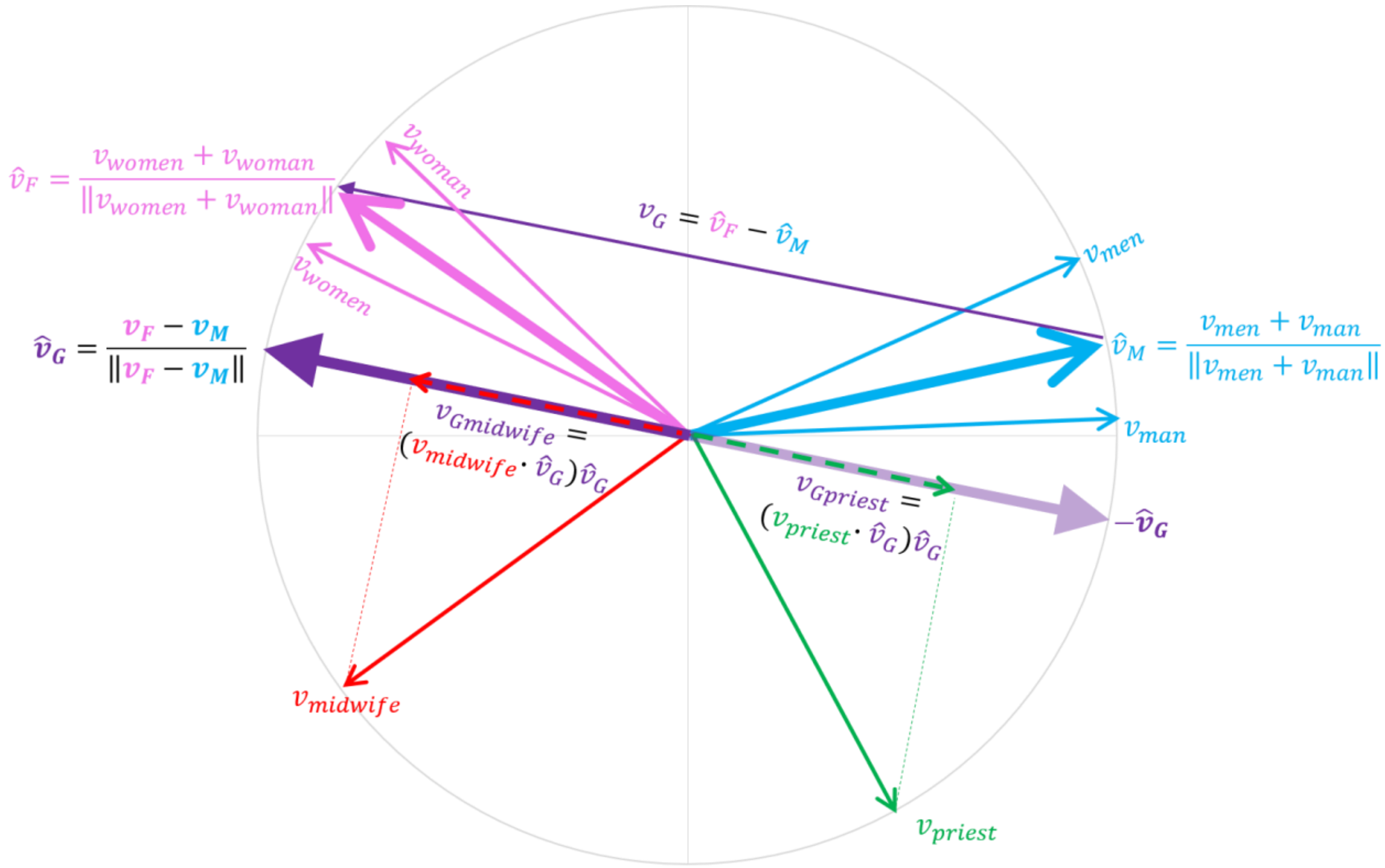

**Figure 9. By aggregating related terms, we can create arbitrary constructs representing cultural concepts. In the figure, the vectors representing the terms *man* and *men* are added to create a length normalized male construct, $\hat{v}_M$. A female construct $\hat{v}_F$ can be created similarly. Subtracting $\hat{v}_M$ from $\hat{v}_F$ results in a vector $v_G$ pointing from masculinity towards femininity which when normalized and centered can represent a gender axis $\hat{v}_G$. Any term ($v_{priest}$) in the model vocabulary can be projected onto this axis, $v_{Gpriest} = (v_{priest} \cdot \hat{v}_G)\hat{v}_G$, to estimate the degree of association of the term with males or females in the corpus on which the model was trained.**

Vector projections on popular word embeddings models

Once a cultural axis, such as gender, has been derived from a word embedding model, terms in the model vocabulary can be projected onto that axis to detect associations in the model between the projected terms and the poles of the axis. Figure 10 shows the results of projecting word vectors denoting professions onto a gender axis estimated from the FastText embedding model trained with subword information on Wikipedia 2017, UMBC webbase corpus and statmt.org news dataset (16B tokens). The landing position of the vector projections on the axis reveals the association of the projected term with the masculine or feminine poles of the axis. These associations are in turned derived from the corpus of textual data on which the word embedding model was trained.

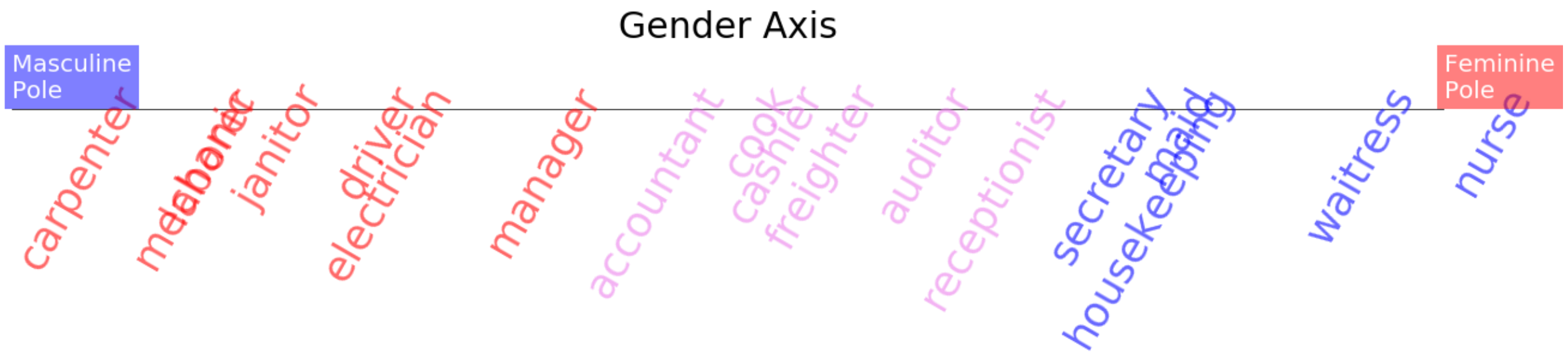

**Figure 10. After creating a gender axis, vectors representing words in the vocabulary of the model can be projected onto the gender axis. This figure shows the projection of words denoting professions onto a gender axis derived from a FastText model (16B tokens).**

<u>Correlation of vector projections with empirical data about the world</u>

Several works have previously shown that in commonly used word embedding models, the value of vector projections on cultural axes or aggregates of related words correlate significantly with quantitative metrics about the empirical world *(5, 7, 8)*. For instance, the vector projection values of words describing professions onto a gender axis has a strong correlation with the percentage of the workforce that is female in those professions. That is, professions with a large representation of women tend to project to the feminine pole of a gender axis derived from a word embedding model. In contrast, professions with low levels of female participation, tend to project to the opposite masculine pole, see Figure 1. Despite most works in the literature focusing on this type of correlation between professions and gender, many other types of cultural axes can be created, such as for instance, socioeconomic axes, or political orientation axes, see Figure 1 and Figure 11.

<u>Data sources used with quantitative metrics about the empirical world</u>

The data sources containing quantitative information about the world used in Figure 1 to demonstrate their association with the geometrical structure of embedding models are listed below:

- Employment data: 2015 Current Population Survey of the U.S. Bureau of Labor Statistics https://catalog.data.gov/dataset/current-population-survey-labor-force-statistics
- Countries GDP: CIA world factbook https://www.cia.gov/library/publications/the-world-factbook/
- Car brand manufacturer prices: US News car rankings (for each brand, the average price of the most expensive and cheapest car from the brand was calculated) https://cars.usnews.com/cars-trucks/browse?make=Buick&make=Cadillac&sort=price_desc
- Democrat to Republican campaign contributions ratios: Federal election commission http://verdantlabs.com/politics_of_professions/index.html

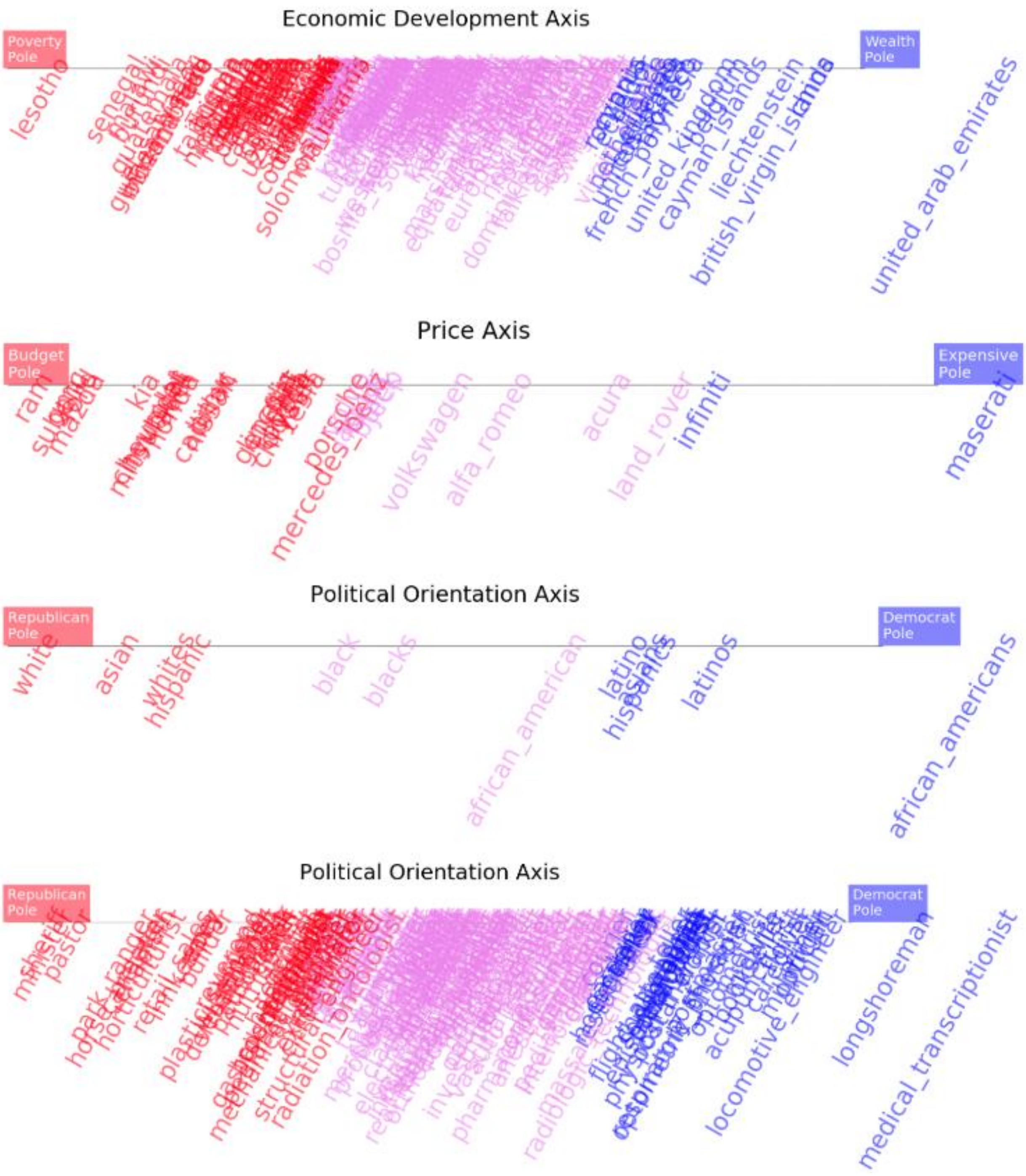

**Figure 11. Cultural axes do not need to be circumscribed to clear cut concepts such as gender. Arbitrary axes describing economic development, socioeconomic status or political orientation can be created. Projecting relevant word vectors onto those axes reveals the associations contained in the corpus (Google News in the Figure) on which the word embedding model (word2vec) was trained.**

Harvard General Inquirer Lexicon

This work has tested whether the association results obtained with the small WEAT lexicon (N=50) in previous scholarly literature (8,16) replicate when using a bigger lexicon of manually labeled terms according to positive and negative polarity. Thus, we use the Harvard General Inquirer (HG) IV-4 (14) positivity/negativity lexicon or HGI for short (N=3623 terms) that has been widely employed in content analysis studies. Note that the HGI contains 4,206 total annotations. Yet, several annotations belong to multiple senses with which a word can be used. For instance, the word *fun* can be used as a noun-adj to indicate enjoyment or enjoyable and has a corresponding positive annotation entry in HGI. But the word *fun* can also be used in the sense of *making fun*, and for that sense there is a corresponding negative annotation entry in HGI. For words with multiple senses, we used the annotation label of the most common usage of the word in the language, in this case positive, since the HGI provides an estimate percentage of sense usage frequency for terms with multiple annotations.

External Lexicons used

This work has used a total of 17 external lexicons, listed in the Appendix, containing terms annotated for positive and negative polarity. The ensemble of sentiment lexicons includes several lexicons often used in the scholarly literature for content and sentiment analysis, several online lists of positive and negative character traits, lists of positive and negative adjectives as well as several specialized lexicons from the General Inquirer that measure constructs with clear positive and negative dichotomies such as vice/virtue, conflict/cooperation or hostility/affiliation. Original lexicons were preprocessed to remove invalid entries such as for instance the emoticons contained in the Vader lexicon since they are not present in the word embeddings models analyzed.

Alignment of cultural axes representing demographic groups with axes derived from Wordnet antonym pairs

To elucidate the nature of the associations between sentiment lexicons and cultural axes, we estimated cultural axes from the 3872 Wordnet antonym pairs. We then calculated the cosine similarity between each one of these axes and the cultural axes representing demographic groups that are the focus of this work. A high degree of cosine similarity indicates alignment of the words in the antonym pair with the poles of the cultural axis, see Figure 12. For example, an axis derived from the antonym pair *maternal-paternal* will have a high degree of alignment (i.e. cosine similarity) with the gender axis. That is, the word *maternal* will be close to the feminine pole of the gender axis formed by words such as *woman*, *women* or *female*. In contrast, the word *paternal* will be close to the masculine pole of the gender axis formed by words such as *man*, *men* or *male*. Thus, both of these axes will be similar in orientation. In contrast, an axis formed by a set of antonyms with no apparent relatedness to neither males nor females, such as *centrifugal-centripetal*, will be more orthogonal (i.e. dissimilar) to the gender axis.

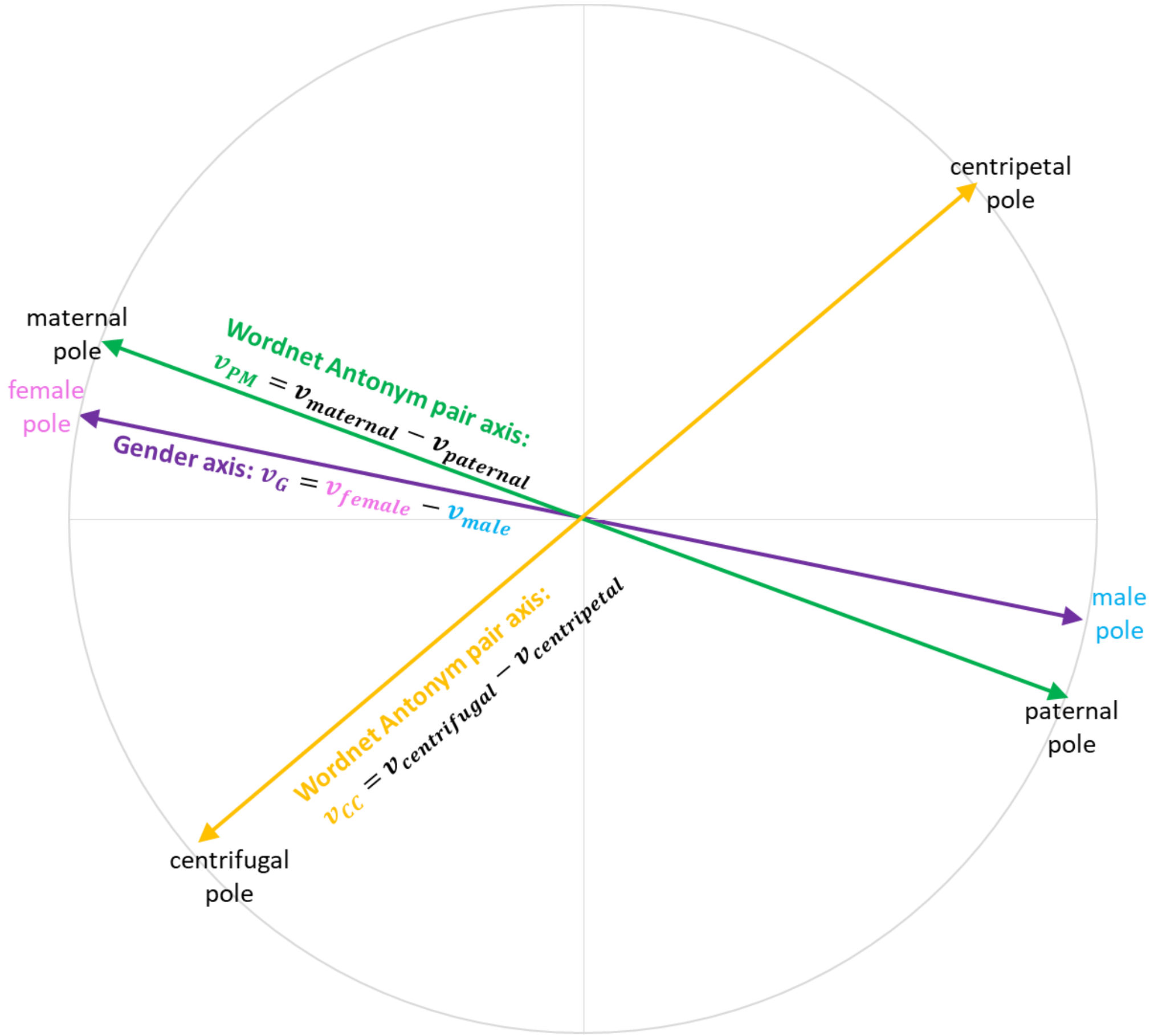

**Figure 12 The cosine similarity between axes created from WordNet antonym pairs and cultural axes representing demographic groups can be estimated to quantify the degree of alignment of the antonym pair with the poles of the cultural axis. An axis formed with words such as maternal and paternal that are obviously related to males and females will have a high degree of alignment with the gender axis. In contrast, antonym pairs with a low degree of relatedness or association with males and females (such as centrifugal-centripetal) will be more orthogonal to the gender axis.**

Parametric versus nonparametric correlation coefficients

All the correlation analyses described in the Results section generate similar outcomes regardless of what correlation coefficient (Pearson or Spearman) is used. Results are reported in this manuscript using the Spearman correlation coefficient since it makes fewer assumptions about the underlying distribution of the data.

GitHub repository description

All the methods and materials necessary to reproduce the results described in this manuscript are available at (https://github.com/drozado/WideRangeScreeningOfAlgorithmicBiasInWordEmbeddings). The folder *analysis* contains all the code and data structures needed to create cultural axes and project lexicons onto them. The folder also contains all the lexicons used in the analyses and the lists of terms used to construct the poles of all the cultural axes. The folder contains scripts as well to evaluate the performance of the seven word embedding models analyzed on metrics such as word pairs similarity, relatedness as well as morphological, lexical, encyclopedic and lexicographic analogies.

The folder *literatureSearch* contains all the query outcomes across 4 bibliographic search engines generated in the search for manuscripts on the topic of biases in word embeddings. The folder *manuscripts* contains the Title page and Introduction section of all manuscripts that passed the selection criteria described above. The manuscripts pdf files have been highlighted to indicate the locations on the text used to justify the classification of manuscripts as citing certain bias types. The folder *tables* contains all the tables displayed in this work and additional metadata.

## Author Contributions Statement

This work was carried out in its entirety by D.R.

## Acknowledgments

NA